\documentclass[runningheads]{svmult}
\usepackage{makeidx}   
\usepackage{graphicx}  
\usepackage{subeqnar}  
\usepackage{multicol}  
\usepackage{cropmark} 
\usepackage{physprbb}  
\begin{document}
\title*{Scalar Field Dark Matter and Galaxy Formation}
\toctitle{Scalar Field Dark Matter and Galaxy Formation}
%
%
\titlerunning{Scalar Field Dark Matter}
%
\author{Miguel Alcubierre\inst{1}
\and F. Siddhartha Guzm\'an\inst{2}
\and Tonatiuh Matos\inst{3}
\and Dar\'{\i}o N\'u\~nez\inst{1,4}
\and L. Arturo Ure\~na-L\'opez\inst{5}
\and Petra Wiederhold\inst{6}}
\authorrunning{Alcubierre, Guzm\'an, Matos {\it et al.}}
%
%
\institute{Instituto de Ciencias Nucleares, Universidad Aut\'onoma de M\'exico,
A.P. 70-543, 04510 D.F., M\'exico.
\and Albert Einstein Institut, Max Planck Institut f\"ur 
Gravitationsphysik, Am M\"uhlenberg 1, D-14476 Golm, Germany
\and Departamento de F\'{\i}sica, Centro de Investigaci\'{o}n y de Estudios
Avanzados del IPN, A.P. 14-740, 07000 M\'{e}xico D.F., M\'exico.
\and Center for Gravitational Physics and Geometry, Penn State University,
University Park, PA 16802, USA.
\and Astronomy Centre, University of Sussex, Brighton BN1 9QJ, United Kingdom.
\and Departamento de Control Autom\'amtico, Centro de Investigaci\'{o}n y de
Estudios Avanzados del IPN, A.P. 14-740, 07000 M\'{e}xico D.F., M\'exico.
}

\maketitle              


\begin{abstract}
We present a general description of the scalar field dark matter
(SFDM) hypothesis in the cosmological context. The scenario of
structure formation under such a hypothesis is based on Jeans
instabilities of fluctuations of the scalar field. It is shown that it
is possible to form stable long lived objects consisting of a wide
range of typical galactic masses around $10^{12}M_{\odot}$ once the
parameters of the effective theory are fixed with the cosmological
constraints. The energy density at the origin of such an object is
smooth as it should.
\end{abstract}


\section{Introduction}

It has been established during this conference that the problem of the
nature of dark matter is a non trivial puzzle which is missing even
candidates that can provide adequate explanations at all relevant
scales\cite{ellis}. Here we push forward the simple hypothesis that
establishes a real scalar field as a candidate to be the dark matter
at both cosmological and galactic scales, and we focus our attention
on the possibility that the structures contained in galaxies can be
formed under such a hypothesis. The scalar field hypothesis is
motivated because the old belief that the whole material content of
the Universe is made of quarks, leptons and gauge bosons is now being
abandoned due to recent observations and
inconsistencies~\cite{schramm}. As it has been suggested, the cure for
the CDM models seems to be the addition of a self-interaction
between the dust particles, which would be automatically included in
the SFDM model just by introducing a potential of self-interaction.\\

The hypothesis that scalar fields could be the dark matter is not
new. For instance, Denhen proposed that the Higgs field could play
that role at galactic scale (see~\cite{denhen}), and other authors
have suggested that galactic halos are in fact boson
stars~\cite{schunck}. Maybe the most popular hypothesis concerning
scalar fields as dark matter is that of the axion stars, which could
account for the missing matter as MACHOS at all scales.  However, it
has been shown that for the possible range of the axion mass such
stars should either fragment or be highly unstable in the time-frame
of gravitational cooling processes, since the axion tends to form
compact objects (oscillatons) in a short dynamical time scale. But,
axions become unstable in such a state and they could be discarded as
candidates~\cite{seidel94}. Nevertheless, when galactic halos are
considered as massive objects made of a scalar field in the same
spirit as axion stars, it has been shown that a single real scalar
field could explain the rotation curves in
galaxies~\cite{flats,flatprd}. It was also possible to present a
solution that could explain the presence of supermassive compact
objects in galactic active nuclei and thus would provide a space-time
background for galaxies in terms of scalar fields, which could explain
the motion of the visible matter in galaxies \cite{franky}. However,
one has to take care of the risks of instability pointed out precisely
when the axion stars where studied, and thus the range of masses for
the scalar field should change dramatically ($10^{18}$ times lighter
than the Peccei-Quinn axion). Such masses
should be restricted by the cosmological scenario.\\

Once the cosmological and galactic scales have been already motivated,
it is now time to show whether or not the parameters fixed with the 
cosmological observations permit the formation of galaxies assuming 
that the dark matter is a scalar field.\\

\section{The Cosmological Model}

Dark energy has been shown to be a fundamental ingredient of any
cosmological model.  Here we will consider dark energy as a
quintessence field consisting of a real scalar field $\Psi$. On the
other hand, our dark matter scalar field would be described by a
second scalar field $\Phi$.  Finally, the ordinary matter of the
Universe (baryons, radiation, neutrinos, etc) is represented by a
perfect fluid field with density $\rho$ and pressure $p$. We assume
the space-time to be FRW, i.e. described by the line element
$ds^2=-dt^2+a(t)^2(dr^2+r^2d\Omega^2)$, thus the equations governing
all these matter components read

\begin{eqnarray}
H^{2} \equiv \left( \frac{\dot{a}}{a} \right)^2 &=&\frac{\kappa _{0}}{3}  
\left( \rho +\rho _{\Phi }+\rho _{\Psi }\right)  \label{fried} \\
\ddot{\Phi}+3H\dot{\Phi}+\frac{dV(\Phi )}{d\Phi } &=&0  \label{cphi} \\  
\ddot{\Psi}+3H\dot{\Psi}+\frac{d\tilde{V}(\Psi )}{d\Psi } &=&0  
\label{cpsi}\\
{\dot{\rho}}+3H\left( \rho +p\right) &=&0,  \label{fluid}
\end{eqnarray}

\noindent where $\kappa _{0} \equiv 8\pi G$. The scalar energy
densities (pressures) are $\rho _{\Phi }=\frac{1}{2}\dot{\Phi}
^{2}+V(\Phi )$ ($p_\Phi =\frac{1}{2}\dot{\Phi}^{2}-V(\Phi )$) and $%
\rho_{\Psi }=\frac{1}{2}\dot{\Psi}^{2}+\tilde{V}(\Psi )$ ($p_\Psi
=\frac{1}{2 }\dot{\Psi}^{2}-\tilde{V}(\Psi )$). Here dots denote
derivative with respect to cosmological time $t$.\\

The SFDM
model~\cite{flats,flatprd,franky,varun,urena,luis2,luis3,luis4}
assumes that both the dark matter and the dark energy are of
scalar field nature. A particular model we have considered in the last
years is the following:  For the dark energy we adopt a quintessence
field ${\Psi}$\ with a scalar potential of the form $\tilde{V}(\Psi )
=\tilde{V_{0}}\left[ \sinh {(\alpha \,\sqrt{\kappa_{0}}\Psi )}\right]
^{\beta }$. It has been shown that this quintessence potential is a
reliable model for the dark energy, because of its asymptotic
behavior~\cite{luis}. Nevertheless, the main point of the SFDM model
is to assume that the dark matter is also a scalar field ${\Phi}$\
endowed with a self-interacting potential (see~\cite{luis2})

\begin{equation}
V(\Phi ) = V_{0}\left[ \cosh {(\lambda \,\sqrt{\kappa _{0}} \Phi 
)}-1\right]
\label{cosh}
\end{equation}

\noindent The mass of the scalar field $\Phi$ is defined as $%
m_{\Phi}^{2}=V^{\prime \prime }|_{\Phi =0}=\lambda ^{2}\kappa
_{0}V_{0}$. This potential allows the scalar field $\Phi$ to
behave just as cold dark matter at cosmological scales, provided that
the mass of the boson particle is given by\\

\begin{equation}
m^2_\Phi = (1.7/3) \, \lambda^2 \left( \lambda^2-4 \right)^3
\Omega^{-3}_{0,\gamma} \Omega^4_{0,{\mathrm CDM}} H^2_0 \label{cmass}
\end{equation}

\noindent where $(\Omega_{0,{\mathrm CDM}},\Omega_{0,\gamma},H_0 )$
are the current values of the density parameters of cold dark matter
and radiation, and the Hubble parameter, respectively. Because of
nucleosynthesis constraints\cite{varun,ferr}, we have to take $\lambda
\geq 5$.

We now analyze the perturbations of the space-time metric due to the
presence of the scalar field $\Phi$. First, we consider a linear
perturbation given by $h_{ij}$. We work in the synchronous gauge
formalism, where the line element is
$ds^{2}=a^{2}[-d\tau^{2}+(\delta_{ij}+h_{ij})dx^{i}dx^{j}]$. We must
add the perturbed equation for the scalar field $\Phi(\tau)
\rightarrow \Phi(\tau) + \phi(k,\tau)$\cite{ferr,stein}

\begin{equation}
\ddot{\phi} + 2 {\cal H}\dot{\phi} + k^2 \phi + a^2 V^{\prime \prime} 
\phi + (1/2) \dot{\Phi} \dot{h} = 0  \label{pphi}
\end{equation}

\noindent to the linearly perturbed Einstein equations in $k$-space
($\vec{k} = k \hat{k}$). Here, dots are derivatives with respect to
the conformal time $\tau$, primes are derivatives with respect to the
unperturbed scalar field $\Phi$, $h$ is the trace of the metric
perturbations $h_{ij}$ and $\cal H$ is the conformal Hubble factor.\\

\noindent It is well known that scalar perturbations can only grow if
the $k^{2}$-term in eq.~(\ref{pphi}) is subdominant with respect to
the second derivative of the scalar potential, that is, if $k<a
\sqrt{V^{\prime \prime}} $ \cite{ma2}. According to the cosmological
solution for potential (\ref{cosh}), the scalar wave number defined by
$k_{\Phi }=a \sqrt{V^{\prime \prime }}$ has a minimum value given by
\cite{luis3}

\begin{equation}
k_{min,\Phi} = 1.3 \, \lambda \sqrt{\lambda^2 -4} \,
\Omega^{-1/2}_{0,\gamma} \Omega_{0,{\mathrm CDM}} H_0 , \label{kmin}
\end{equation}

\noindent It can be assured that there are no scalar perturbations for
$k > m_\Phi$, that is, bigger than $k_\Phi$ today: they must have been
completely erased. Besides, modes such that $m_\Phi > k >
k_{min,\Phi}$ must have been damped during certain epochs.
From this, we conclude that the scalar power spectrum of
$\Phi$ will be damped for $k > k_{min,\Phi}$ with respect to the
standard case. Therefore, the scalar Jeans length associated to
potential~(\ref{cosh}) must be \cite{luis3}

\begin{equation}
L_J (a) = 2 \pi \, k^{-1}_{min,\Phi}.  \label{jl}
\end{equation}

In Fig. 1 we can see the damped density contrast $\delta ^{2}$ at a
redshift $z=50 $ from a complete numerical evolution using {\small
CMBFAST}.  We see that it is possible to suppress the power spectrum
in such a manner that it is possible to 
explain the smooth cores of dark halos in galaxies and the small
abundance of dwarf galaxies, as suggested in~\cite{kamion}. 
The mass power spectrum is related to
the CDM case by the semi-analytical relation (see \cite{hu})

\begin{figure}[b]
\begin{center}
\includegraphics[width=.5\textwidth]{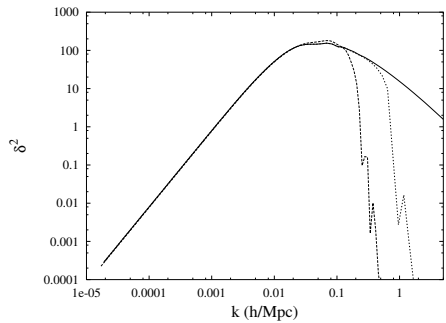}
\end{center}
\caption[]{Power spectrum at a redshift $z=50$: $\Lambda$CDM (solid-curve),
and $\Phi $CDM with $\protect\lambda = 5$ (dashed-curve) and $\protect%
\lambda = 10$ (dotted-curve). The normalization is arbitrary.}
\label{eps1.1}
\end{figure}

\begin{equation}
P_\Phi(k) \simeq \left( \frac{\cos{x^3}}{1+x^8} \right)^2 P_{CDM}(k),
\end{equation}

\noindent but using $x=(k/k_{min,\Phi})$, with $k_{min,\Phi}$ being the 
wave number associated to the Jeans length (\ref{jl}). If we take a 
cut-off of the mass power spectrum at $k=4.5 \, h \, {\rm Mpc}^{-1}$ 
\cite{kamion}, we can fix the value of the parameter $\lambda$. Using eq. 
(\ref{kmin}) we find that \cite{luis3}

\begin{eqnarray}
\lambda &\simeq& 20.3,  \nonumber \\
V_0 &\simeq& \left(3.0\times 10^{-27}\,M_{Pl}\simeq 36.5\, {\rm eV}\right)
^{4}, \label{parametros}\\
m_\Phi &\simeq& 9.1\times 10^{-52}\,M_{Pl}\simeq 1.1\times 10^{-23}\, {\rm 
eV}.  \nonumber
\end{eqnarray}

\noindent where $M_{Pl}=1.22 \times 10^{19} \, {\rm GeV}$ is the Plank 
mass. All parameters of potential (\ref{cosh}) have now been completely 
determined and we have the right cut-off in the mass power spectrum.\\

\section{Dark Matter in Galaxies}

The problem of galactic dark matter is directly related to the
rotation curves of visible matter, for which we started by proposing a
model for the dark matter dominated
region~\cite{flats,flatprd}. There, it was assumed that the scalar
field was the only contributor to the energy density in that region
and thus the one determining the structure of the space-time there;
some properties like a relation $p_r = 10^6 \rho$ between the radial
pressure of the scalar fluid and the energy density of the field were
discovered \cite{flatprd}. Nevertheless, it remained unclear if there
would be a source that could support such scalar field, since it is
well known that there are no non-singular solutions for spherically
symmetric static scalar field configurations. We showed later, through
a toy model made of scalar fields, that a ball of dust and an frozen
oscillaton could be the source for such a scalar field halo and in one
step we found the structure of the space-time from the galactic center
to the dark matter dominated region \cite{franky}. But probably the
major result is the suggested line element found for such a toy model

\begin{equation}
ds^{2}=-B_{0}(r^{2}+b^{2})^{v_{a}^{2}}\left( 1-\frac{2M}{r}\right) 
dt^{2}+\frac{A_{0}}{(1-\frac{2M}{r})}dr^{2}+r^{2}d\Omega ^{2}  
\label{met_bosones}
\end{equation}

\noindent  with $d\Omega ^{2}=d\theta ^{2}+\sin ^{2}\theta d\varphi ^{2}$ 
and $M$ a constant related to the mass of the central supermassive 
object in a way discussed in ref \cite{franky}. $B_0$, $b$ and $A_0$ are 
integration constants and $v_a$ is the asymptotic value of the velocity 
of test particles in the sense of the Tully-Fisher relation. This metric 
is singular at $r=0$, but it has an event horizon at $r=2M$. This metric 
does not represent a black hole because it is not asymptotically flat.
Nevertheless, for regions where $r\ll b$ but $r>2M$\ the metric behaves 
like a Schwarzschild black hole. Inside of the horizon the pressure of the
perfect fluid is not zero, therefore it does not behave as dust anymore. 
Thus our toy model is valid only in regions outside the horizon, where 
it could be an approximation of the galaxy. Metric (\ref{met_bosones}) is 
not asymptotically flat, but it has a natural cut off when the dark matter 
density equals the intergalactic density as mentioned in 
\cite{flatprd}.\newline

The charm of such a metric consists in correctly reproducing the
rotation curves of galaxies from the center of the galaxy to the dark
matter dominated region. The expression for the tangential velocity
$v^{rot}$ of test particles moving along circular stable orbits, as
seen by an observed at infinity, is given by
$v^{rot}=\sqrt{rg_{tt}{}_{,r}/(2g_{tt})}$ for any static spherically
symmetric space-time (see \cite{flatprd}). Thus for
(\ref{met_bosones}) this reads

\begin{equation}
v^{rot}(r)= \sqrt{\frac{v^{2}_{a}(r-2M)r^2
+M(r^2+b^{2})}{(r-2M)(r^2+b^{2})}},
\label{velprof}
\end{equation}

\noindent a formula that allows one to fit observational
curves. Observe that, for $M=0$, the velocity profile (\ref{velprof})
corresponds exactly to the pseudo-isothermal velocity profile for
rotation curves \cite{Rubin}. In Figure 2 we show the variation of the
rotation curve when the value of $M$ changes.  It is evident that this
would affect only the kinematics in the central parts of the galaxy,
exactly in the same way as the mass of the central object should do
\cite{diego}. In Fig. 3 we present the effect on the rotation curves
for test particles traveling on such space-time caused by a change of
the parameter $b$ (a sort of core radius).  Here one can see different
shapes only at large radii, precisely where the dark matter starts to
be important. Therefore, considering the visible matter as test
particles traveling on such a background space-time the scalar field
dark matter in halos seems to be promising.\\

\begin{figure*}[b]
\includegraphics[width=.3\textwidth]{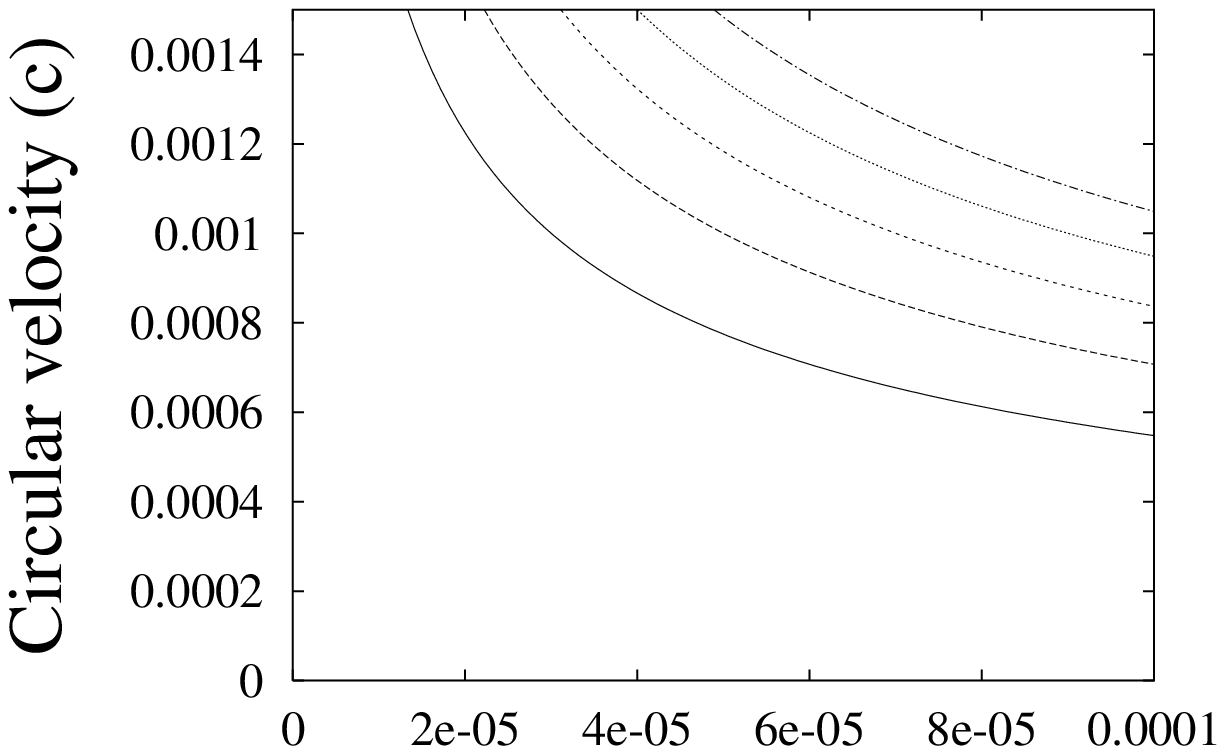}
~\hspace{-1cm}\includegraphics[width=.3\textwidth]{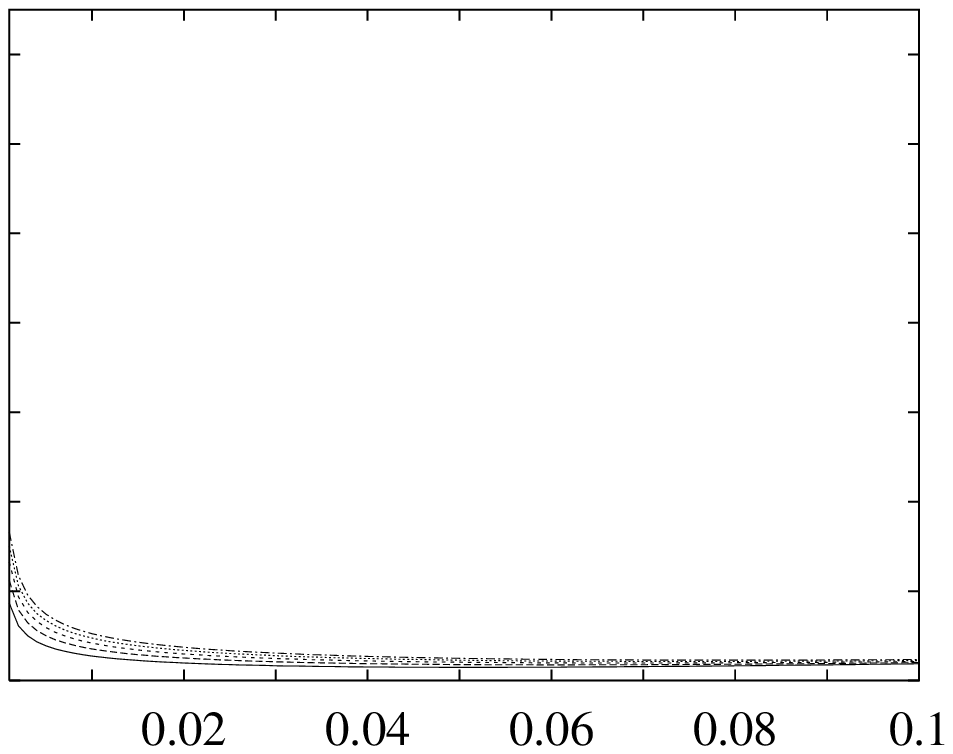}
~\hspace{-1cm}\includegraphics[width=.3\textwidth]{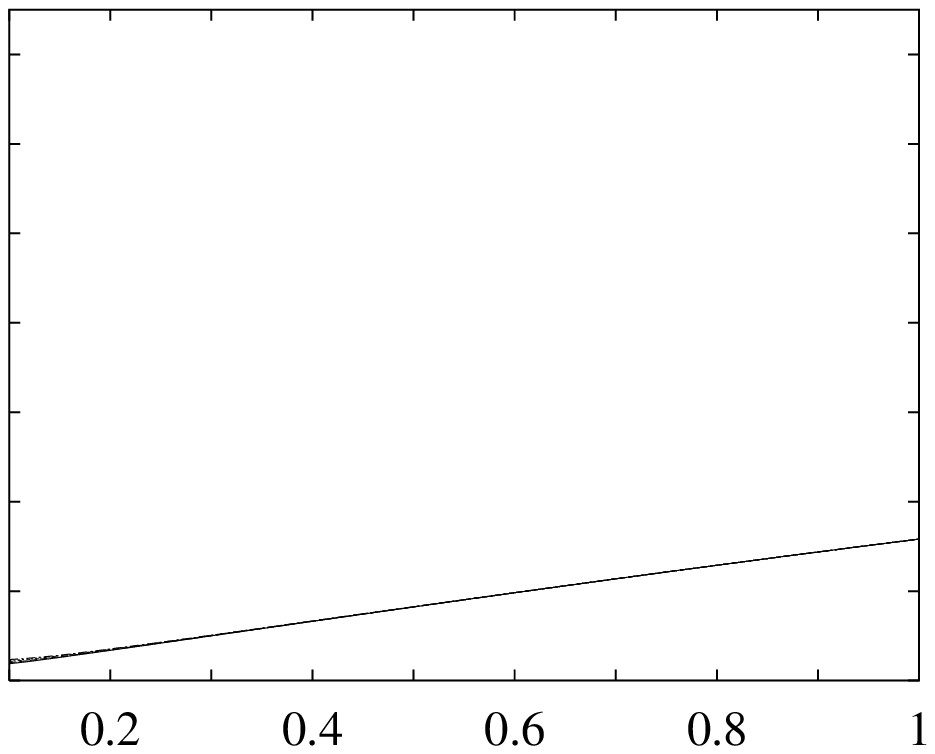}
~\hspace{-1cm}\includegraphics[width=.3\textwidth]{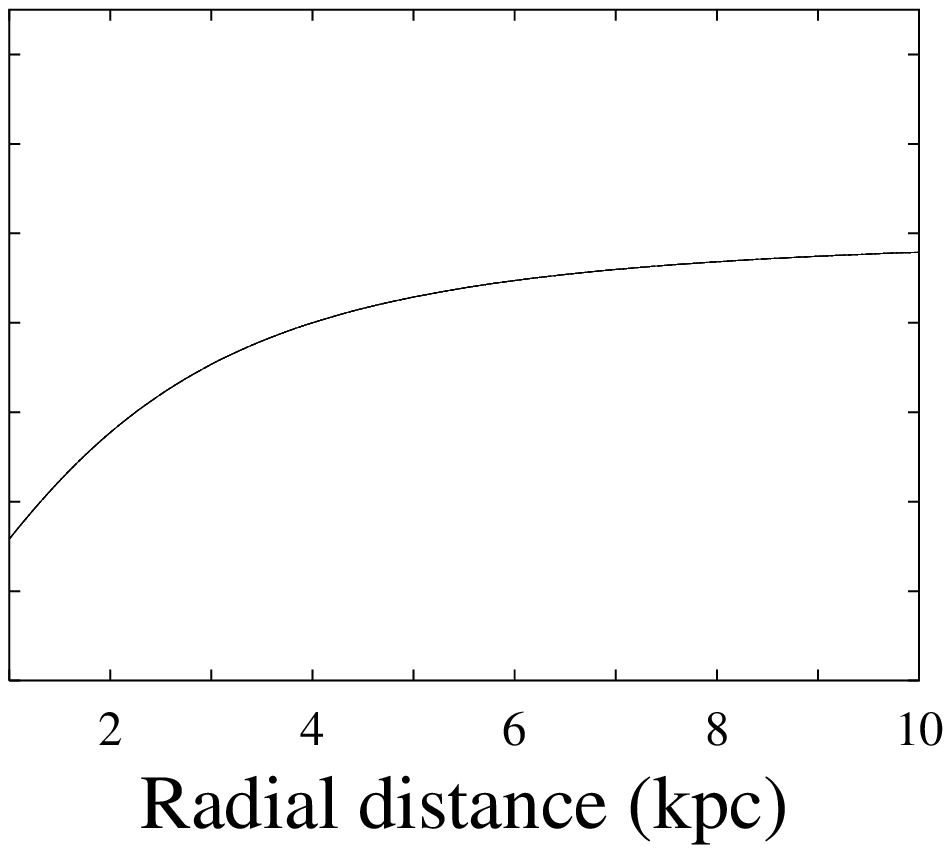}
\caption[]{Rotation curves provided by the line element (10) for 
different values of the mass of the central object: $M=3,5,7,9,11 \times
10^{6}M_{\odot}$ and fixed $(b,v_a)=(3kpc,10^{-3}c)$ are shown.
In the first two plots the rotation curve in the region near the galactic
center, where a keplerian fall off can be observed. In the last two plots
we show the region where the rotation curve associated
to the dark matter component is evident.}
\end{figure*}

\begin{figure*}[b]
\includegraphics[width=.3\textwidth]{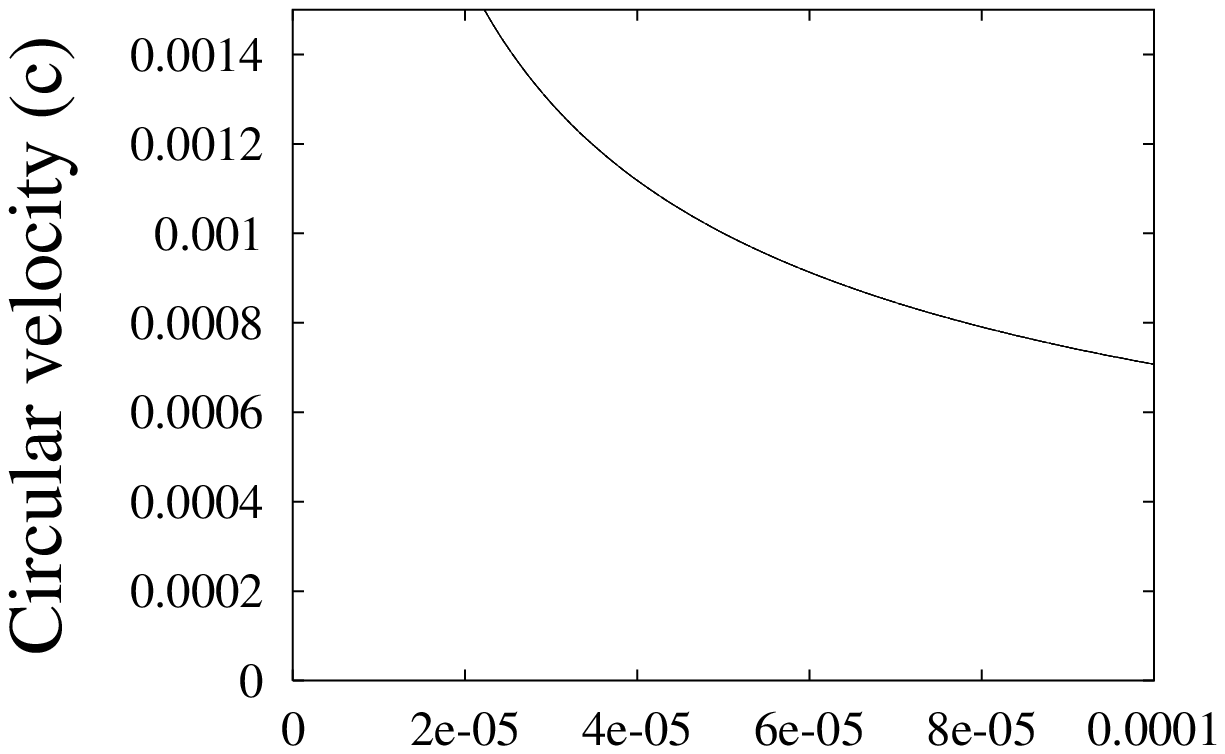}
~\hspace{-1cm}\includegraphics[width=.3\textwidth]{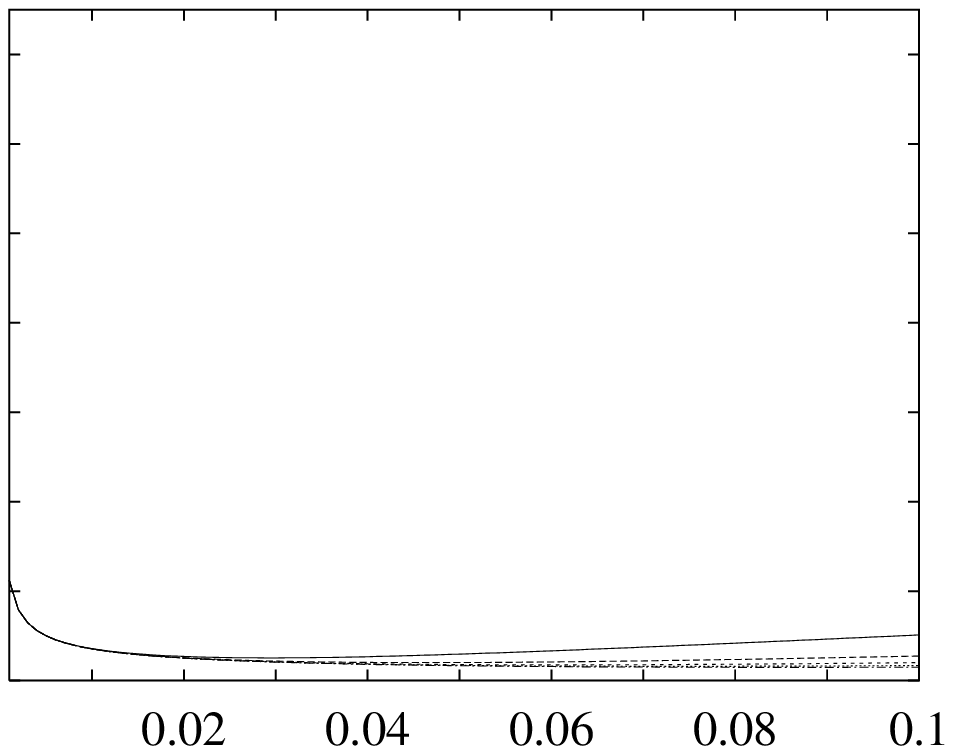}
~\hspace{-1cm}\includegraphics[width=.3\textwidth]{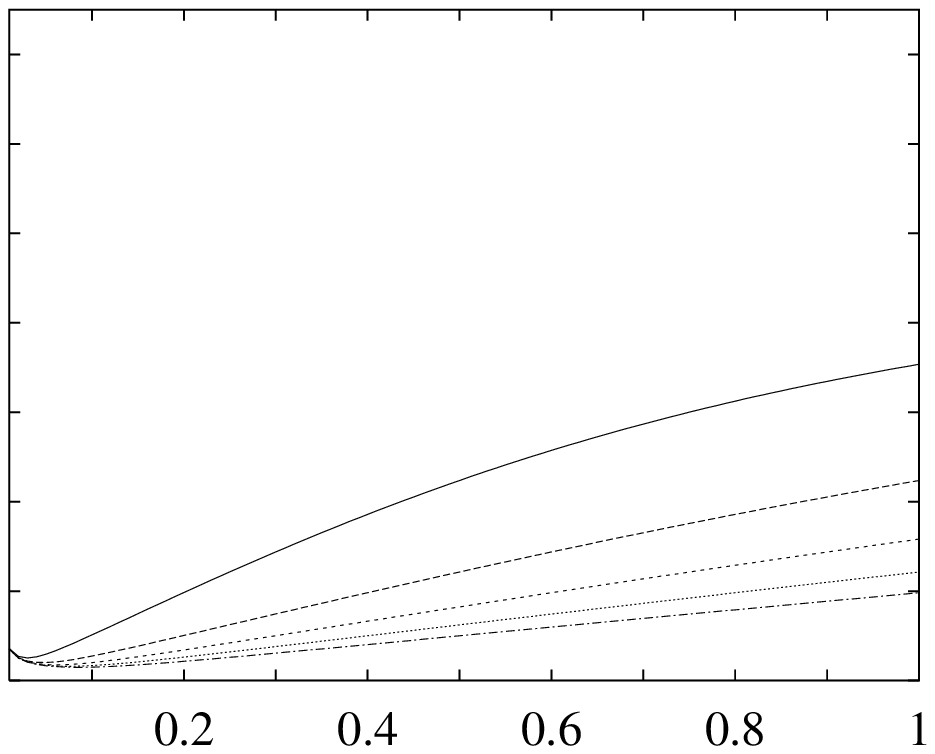}
~\hspace{-1cm}\includegraphics[width=.3\textwidth]{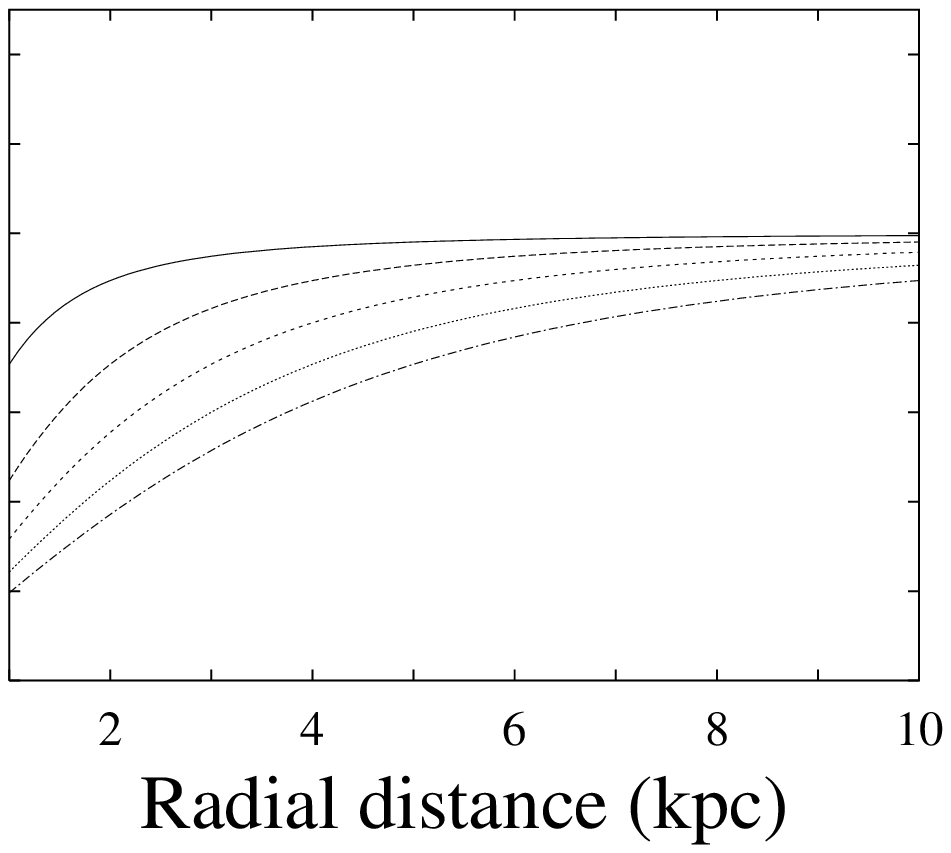}
\caption[]{We show the rotation curves for different values of the
parameter $b=1,2,3,4,5$kpc and with $(M,v_a)=(5 \times
10^{6}M_{\odot},10^{-3}c)$ now fixed. It is evident that in such a
case the shape of the curve in the region associated with the dark
matter changes, but in the center the curves remain unchanged.}
\end{figure*}

\section{Collapse of Scalar Field Dark Matter due to Jeans Instabilities}

In order to study the galactic collapse it was necessary to consider a
spherically symmetric space-time whose components where time
dependent, in such a way that it would be possible to follow the
evolution of the system:

\begin{equation}
ds^2 = -\alpha^2(r,t) dt^2 + a^2(r,t) dr^2 + r^2 d\Omega^2
\end{equation}

\noindent using usual spherical coordinates. To simulate a Jeans
instability we started by considering the initial scalar field to have
a Gaussian distribution $\sqrt{\kappa_{0}}\, \Phi (x,t=0) = A
e^{-x^{2}/s^{2}}$.  The initial mass of the system would change by
varying the width and amplitude of such a Gaussian pulse. The energy
density is given by

\begin{equation}
\rho_s =\frac{1}{2a^{2}} \left( \Phi_{,r} ^{2}+a^2\Phi_{,t}^{2}/\alpha^2 
\right) + V 
\label{densidad}
\end{equation}

\noindent with $V$ given by (\ref{cosh}). According to the results found 
in \cite{seidel94} the critical mass to have configurations that collapse 
are found at $M_{crit}\sim 0.6m_{Pl}^{2}/m \sim 10^{-6}M_{\odot}$ when the 
mass of the boson is $m\sim 10^{-5}eV$ (the case of the axion). But in our 
case the mass of the boson is given by

\begin{equation}
M_{crit}\simeq 0.1 \frac{m_{Pl}^{2}}{\sqrt{\kappa_0 V_0}} =
2.5 \times 10^{13}M_{\odot} \, .
\label{masa1}
\end{equation}

\noindent for the values of the parameters given in~(\ref{parametros}).\\

\begin{figure*}[b]
\begin{center}
\includegraphics[width=.4\textwidth]{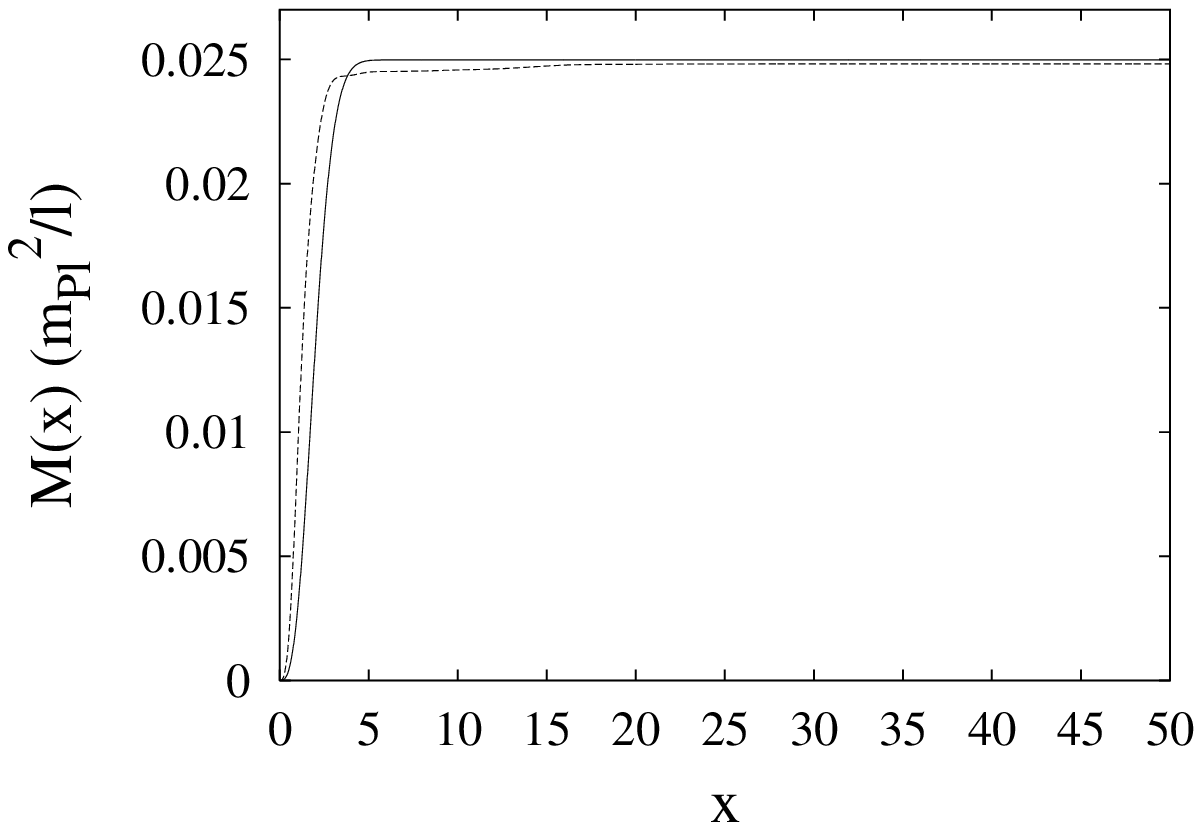}
\includegraphics[width=.4\textwidth]{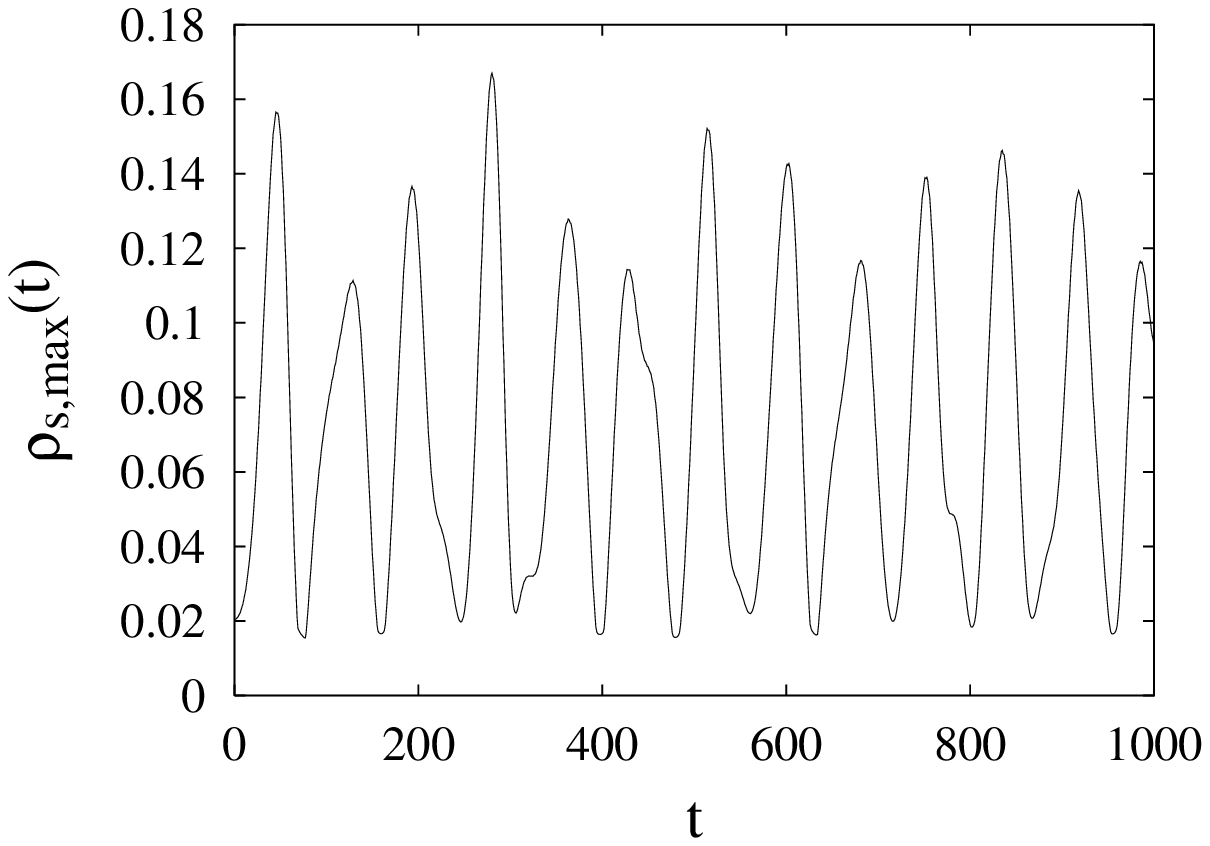}
\includegraphics[width=.4\textwidth]{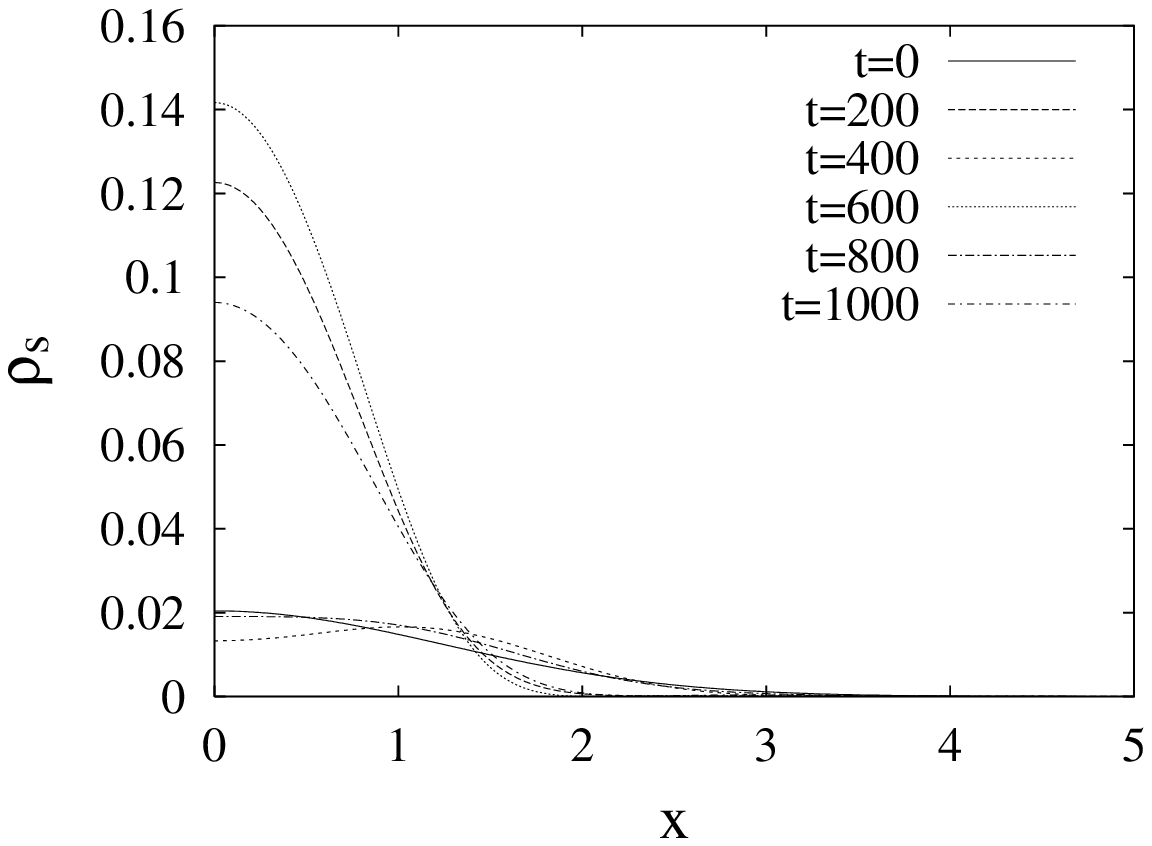}
\end{center}
\caption[]{Left: Integrated mass of the system in terms of time. Right: 
  Temporal evolution of the maximum value of the energy
  density. The parameters for the initial configuration are $A=0.01,\,
  s=2.5$, which correspond to an initial mass of
  $M_i = 6.36 \times 10^{12} \, M_{\odot}$. Center (bottom): Snapshots of 
  the density at different time steps, which shows its smoothness at the 
  origin.} 
\end{figure*}

In Fig. 4 we show the total integrated mass of the system, defined as
$M(x)=\frac{1}{2}\frac{m_{Pl}^{2}}{l}\int_{0}^{x}\rho_{s}(X)\,X^{2}dX$,
as it evolves in time ($x=lr$, $l=\sqrt{\kappa_0 V_0}$).  A small drop
of $\sim 0.6\%$ in the total integrated mass can be observed, but
numerical convergence tests suggest that most (if not all) of this
mass loss is caused by a small amount of numerical dissipation still
present in our numerical method.  This implies that the system does
not radiate any significant amount of energy during the time of the
simulation, which indicates that the object is very stable.  We also
show the value of the central density evolving in time. From this plot
it is evident that the object is oscillating
\cite{pre_galaxia}. Finally, we show the energy density of the scalar
field at different times of the evolution.  The energy density is
clearly smooth at the origin. The boundaries of the object are located
at $x=50$ and the evolution was carried out till $t=1000$, which
corresponds to some 20 light crossing times through the system.\\

\section{Concluding remarks}

The importance of the value fixed for
$m_{\Phi}$~(\ref{cmass},\ref{parametros}), i.e. of the combination of
$\lambda$ and $V_0$, consists precisely in determining the final stage
of the system: if $\lambda$ and $V_0$ are much bigger it is easy to
show that a system of around $10^{12}M_{\odot}$ would collapse into a
black hole, and if they were much smaller the system would simply
disperse. Thus, it is important to stress that we have shown that
precisely the values fixed by the cosmology are those that provide
stable fully relativistic oscillatons with masses around the observed
galactic ones.\\

On the other hand, it could be thought that once the numbers $\lambda$
and $V_0$ are fixed, which means that the effective theory coupling
the scalar field to gravity has all its parameters fixed, the
resulting galaxies should be all of the same mass, but this is not the
case. In fact, we have found that there is a window of combinations of
the initial parameters $A$ and $s$ that allows the formation of
stable objects with different masses. The case presented in Fig. 4
corresponds to a stable and long lived system whose mass is around
twice that of the object presented in~\cite{galaxia}, showing
that the galaxies formed through such a process must not have
the same mass.\\

Moreover, the energy density of the collapsed object is smooth, as
seen also in Fig. 4, which completes the set of nice properties of the
Scalar Field Dark Matter model.\\


\noindent{\bf Acknowledgements}\\

TM would like to thank the organizers for the kind invitation to this conference.
This work was partly supported by CONACyT, M\'exico under gants 34407-E, 32138-E,
010106 (FSG) and 010385 (LAU). DN acknowledges CONACyT and DGAPA-UNAM 
grants for partial support. We thank Ed Seidel for useful discussions. 
We want to thank Aurelio Espiritu and Erasmo Gomez for thechnical support.


%

\end{document}